\documentclass[aip,twocolumn,10pt]{revtex4}%
\usepackage{amsmath}
\usepackage{amsfonts}
\usepackage{amssymb}
\usepackage{graphics}
\usepackage{graphicx}
\providecommand{\U}[1]{\protect\rule{.1in}{.1in}}

\begin{document}
\title{Gyrator Operation Using Josephson Mixers}
\author{Baleegh Abdo}
\author{Markus Brink}
\author{Jerry M. Chow}
\affiliation{IBM T. J. Watson Research Center, Yorktown Heights, New York 10598, USA.}
\date{\today }

\begin{abstract}	
 Nonreciprocal microwave devices, such as circulators, are useful in routing quantum signals in quantum networks and protecting quantum systems against noise coming from the detection chain. However, commercial, cryogenic circulators, now in use, are unsuitable for scalable superconducting quantum architectures due to their appreciable size, loss, and inherent magnetic field. We report on the measurement of a key nonreciprocal element, i.e., the gyrator, which can be used to realize a circulator. Unlike state-of-the-art gyrators, which use a magneto-optic effect to induce a phase shift of $\pi$ between transmitted signals in opposite directions, our device uses the phase nonreciprocity of a Josephson-based three-wave-mixing device. By coupling two of these mixers and operating them in noiseless frequency-conversion mode, we show that the device acts as a nonreciprocal phase shifter whose phase shift is controlled by the phase difference of the microwave tones driving the mixers. Such a device could be used to realize a lossless, on-chip, superconducting circulator suitable for quantum-information-processing applications. 
\end{abstract}

\maketitle

\newpage

\section{Introduction}
Performing high-fidelity, quantum nondemolition measurements in the microwave domain is an important requirement for operating a superconducting quantum computer. Such a requirement is enabled, in various schemes, by using nonreciprocal microwave devices having asymmetrical transmission through their ports, such as circulators \cite{Pozar,Collin,NoiselessCirc,ReconfJJCircAmpl,circulatorLehnert,NRAumentado1,NRAumentado2,circulatorDiVincenzo1,circulatorDiVincenzo2} and low-noise, directional amplifiers \cite{ReconfJJCircAmpl,DircJPC,JDA,JTWPA,KIT,TWPAthreewavemix}. Circulators in particular play several crucial roles in these measurement schemes. They (1) allow quantum systems to be measured in reflection, (2) enable the use of reflective Josephson parametric amplifiers (JPAs) as a first-stage amplifier in the output chain, (3) protect quantum systems against noise in the measurement chain, such as amplified signals reflected off JPAs, excess backaction of directional amplifiers, strong microwave tones feeding JPAs, and noise coming from higher amplification stages of the output line. However, state-of-the-art commercial, cryogenic circulators, ubiquitously used throughout the field today, limit the maximum achievable measurement fidelity and scalability of quantum processors, due to their insertion loss $\sim0.5$ dB, large size $\sim28$ $\operatorname{cm^3}$, and weight $\sim41$ $\operatorname{g}$ \cite{quinstar}. They can also potentially negatively affect the coherence times of qubits, as they are comprised of bulk materials that thermalize poorly and employ strong inherent magnetic fields $10-100$ $\operatorname{G}$ \cite{quinstar}.

\begin{figure}
	[tb]
	\begin{center}
		\includegraphics[
		width=\columnwidth 
		]%
		{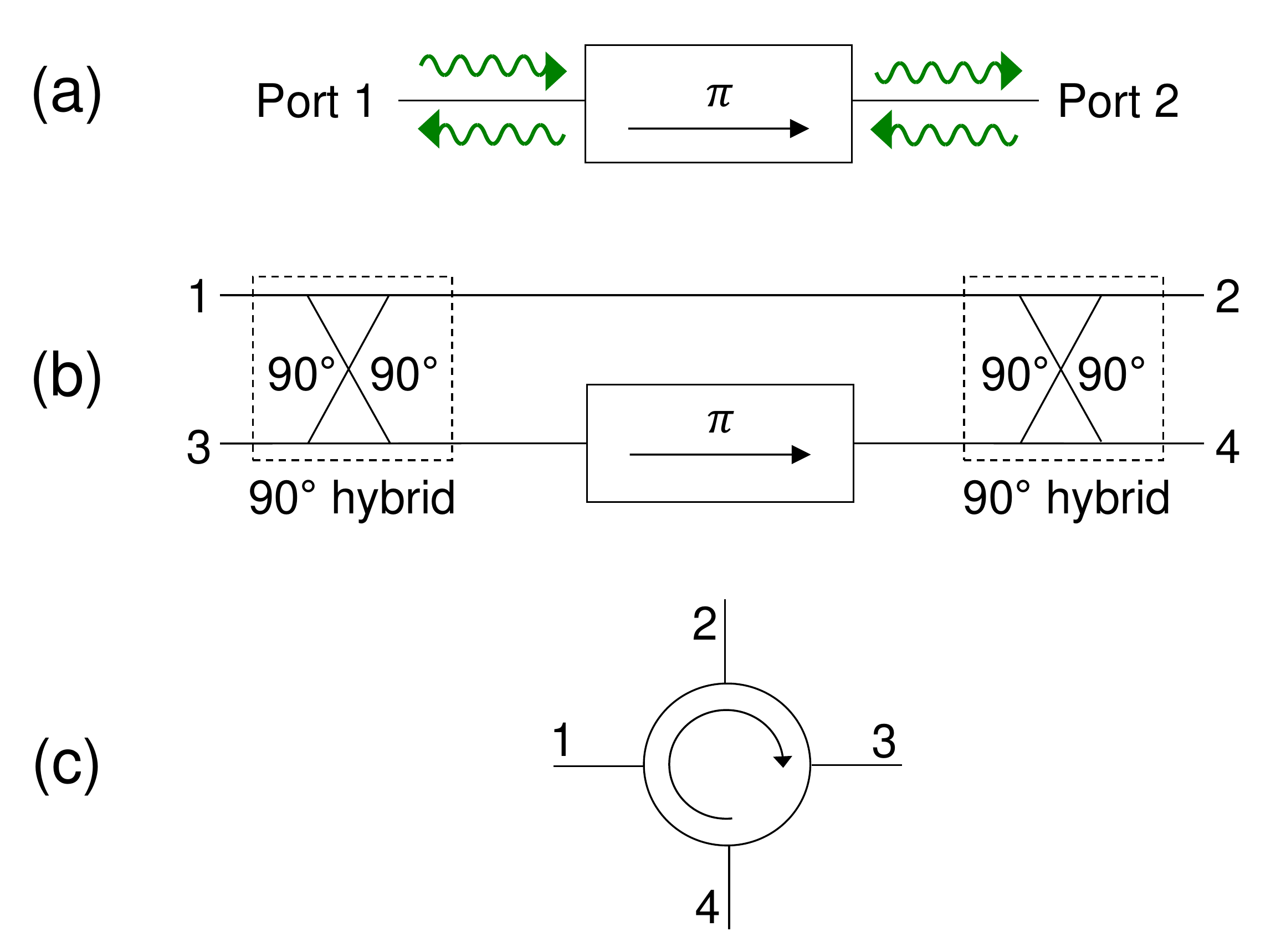}
		\caption{(a) Gyrator symbol. Signals that transverse the device in the direction of the arrow acquire a differential phase shift of $180^{\circ}$ relative to signals propagating in the opposite direction. (b) A four-port circulator can be built by integrating a gyrator into one arm of a Mach-Zehnder interferometer formed using two $90^{\circ}$ hybrids (which function as $50:50$ beams pitters for microwaves). (c) Symbol for the four-port circulator shown in (b). The direction of the arrow defines the circulation direction, i.e., signals entering port $i=1,2,3,4$ are fully transmitted to port $j=2,3,4,1$ respectively. In other words, port $i$ is isolated from signals input on port $j$.      
		}
		\label{gyratorAndcirculator}
	\end{center}
\end{figure}

Nonreciprocal devices, which violate Lorentz reciprocity, break the invariance of the transmission parameter upon exchanging the source and detector \cite{Pozar,Collin,ReciprocityUnitarityTimeReversal}. In this work, we demonstrate a successful operation of an important two-port, nonreciprocal circuit element, i.e., a gyrator, which can be used to build a circulator. A gyrator is a two-port, unity-transmission device, which introduces a differential phase shift of $180^{\circ}$ to microwave signals transversing the device in opposite directions \cite{Pozar,Collin}, as shown in Fig.\ref{gyratorAndcirculator} (a). The scattering matrix of an ideal gyrator reads,
\begin{align}
\left[  S\right]    & =\left(
\begin{array}
[c]{cccc}%
0 & 1 \\
-1 & 0 
\end{array}
\right).  \label{S_gyr}%
\end{align}

By incorporating an on-chip, lossless gyrator into one arm of a Mach-Zehnder interferometer, employing two hybrids \cite{Pozar}, as shown in Fig. \ref{gyratorAndcirculator} (b), it is straightforward to obtain a four-port, on-chip, lossless circulator [Fig. \ref{gyratorAndcirculator} (c)] suitable for scalable quantum processors. In the circulator scheme shown in Fig. \ref{gyratorAndcirculator} (b), circulation is achieved via the generation of two-path interference for waves propagating through the device. The two-path interference is enabled by the addition of two $90^{\circ}$ hybrids, one on either side of the gyrator, which act as $50:50$ beam splitters for microwaves. For example, signals entering port 1 are fully transmitted to port 2 because half of the signal passing through the top arm of the interferometer constructively interferes with the other half of the signal passing through the gyrator incorporated into the bottom arm. On the other hand, signals entering port 2 would destructively (constructively) interfere on port 1 (3), resulting in isolation for port 1 and transmission to port 3. 

In contrast to standard gyrator realizations, which utilize a magneto-optic Faraday effect in order to induce a nonreciprocal phase shift between microwave beams propagating in opposite directions, our scheme is based on a different physics. It utilizes the nonreciprocity of a photonic Aharonov-Bohm effect \cite{AhranovBohmPhotonic} that is analogous to the original effect associated with charged particles \cite{AharonovBohm}. While, in the electronic Aharonov-Bohm effect, electrons propagating in the presence of a magnetic vector potential acquire a nonreciprocal phase that depends on the direction of propagation, in the photonic counterpart, a nonreciprocal phase is acquired by photons propagating in the presence of an artificial gauge field. Such artificial gauge fields for photons can be generated by parametrically modulating certain properties of the photonic system, such as the dielectric constant of dielectric slab waveguides, which induce photonic transitions between two modes \cite{AhranovBohmPhotonic}; the resistance of microwave mixers, which convert the frequency of transmitted signals \cite{AhranovBohmMixers}; and the refractive index in  rf-controlled phase modulators, which shift the phase of propagating optical signals \cite{nonreciprocityPhaseModulators}. In general, the artificial gauge potential in such dynamical schemes corresponds to the modulation phase of the drive (pump). In our case, we generate a similar gauge field for propagating microwave signals by controlling the phase of a microwave drive, which parametrically modulates the dispersive inductive coupling between two nondegenerate resonance modes in dissipationless, three-wave mixers. As a result of this fundamental difference in the device physics, the main advantage of our proposed gyrator scheme over standard realizations is that it employs neither magnetic materials nor strong permanent magnets for its operation. Therefore, it can be low loss, integrated on chip, and fully compatible with superconducting circuits for extensible architectures.
  
\begin{figure*}
	[tb]
	\begin{center}
		\includegraphics[
		width=1.5\columnwidth 
		]%
		{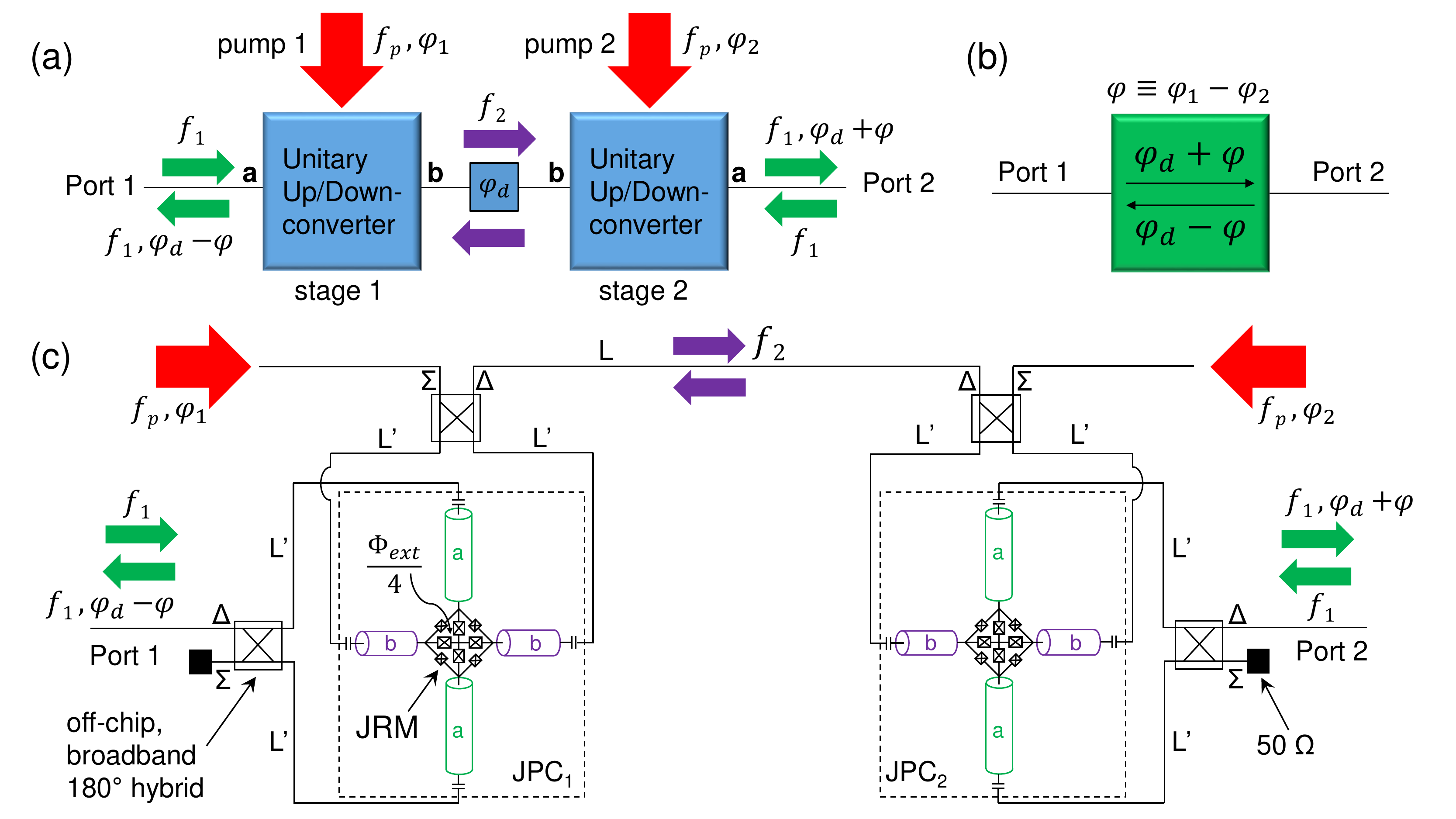}
		\caption{(a) A nonreciprocal phase-shifter scheme, which comprises of two coupled, nondegenerate three-wave mixing devices capable of performing unitary frequency conversion. Each three-wave mixing device has two ports \textit{a} and \textit{b}, which  support incoming and outgoing signals at frequencies $f_1$ and $f_2$, respectively (where $f_2>f_1$). In the exemplary scheme shown in panel (a), the two devices are coupled through port \textit{b}. The device transmits with frequency-conversion input signals at $f_1$ ($f_2$) on port \textit{a} (\textit{b}) to output signals on port \textit{b} (\textit{a}) at $f_2$ ($f_1$). The frequency-conversion process is enabled through energy exchange with the pump drive, feeding the three-wave-mixing device at frequency $f_p=f_2-f_1$. The nonreciprocal phase shift imprinted on the transmitted signals through the device (from port 1 to 2 and vice versa) is set by the phase difference between the two pump drives feeding the two three-wave mixing stages $\varphi$ and the electrical delay between them (i.e., the phase shift $\varphi_d$). (b) A black-box representation of the nonreciprocal phase shifter. Signals transmitted through the device acquire a phase shift of $\varphi_d+\varphi$ from port 1 to 2 versus $\varphi_d-\varphi$ in the opposite direction. (c) A circuit diagram of the proof-of-principle nonreciprocal phase-shifter device, realized by coupling the idler ports of two nominally identical JPCs operating in frequency-conversion mode. The signal ports of the two JPCs form ports 1 and 2 of the whole device. The black lines in the diagram represent normal-metal coaxial cables of lengths L=4" ($10.16$ $\operatorname{cm}$) and L'=2" ($5.08$ $\operatorname{cm}$). The fabrication process of the JPCs is the same as in Ref. \cite{hybridLessJPC}.    
		}
		\label{Device}
	\end{center}
\end{figure*}

While there have been numerous promising schemes which have been shown recently, either experimentally or theoretically, to perform a circulation operation suitable for quantum signals \cite{NoiselessCirc,TRSbreakingInCQED,ReconfJJCircAmpl,circulatorLehnert,NRAumentado1,NRAumentado2,circulatorDiVincenzo1,circulatorDiVincenzo2}, the main two differences between our proposed scheme and those relying on frequency conversion between three modes of a Josephson-based superconducting device, namely, Refs. \cite{ReconfJJCircAmpl} and \cite{NRAumentado2}, are that our proposed circulator scheme preserves the frequency of the input and output quantum signals and requires, in principle, only one pump tone instead of three. This can result in a significant reduction in the overall control hardware resource for operating a larger number of devices. A more detailed comparison is presented in Sec. VII.      
  
  \begin{figure*}
  	[tb]
  	\begin{center}
  		\includegraphics[
  		width=1.5\columnwidth 
  		]%
  		{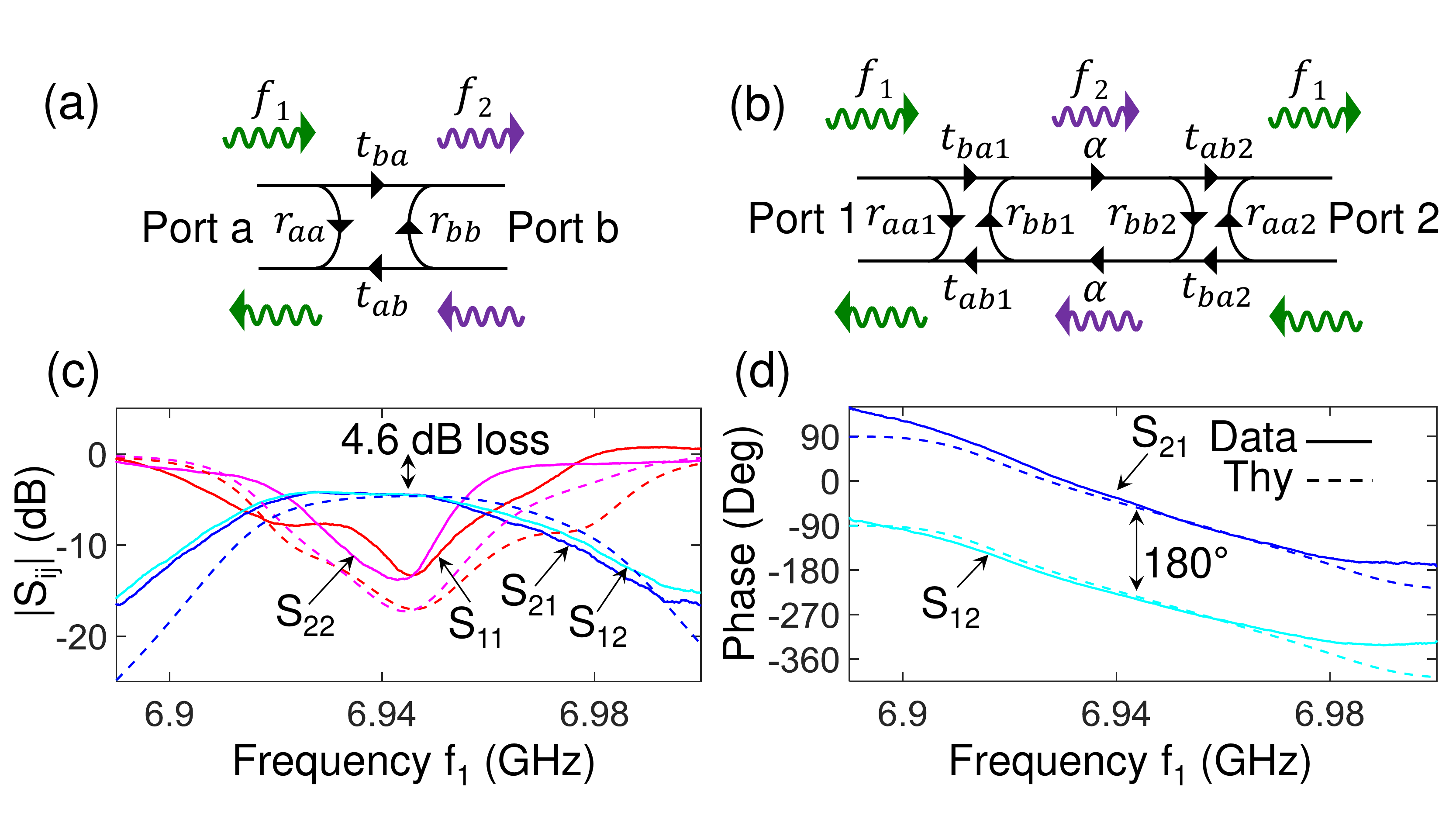}
  		\caption{(a) Signal-flow graph for a JPC device operating in frequency-conversion mode. (b) Signal-flow graph for a two-stage JPC (TSJPC) device. The two JPCs are coupled through their idler ports using a lossy transmission line. (c) Magnitude measurement (solid lines) and calculation (dashed lines) of the scattering parameters of the TSJPC device operated in frequency-conversion mode versus input frequency $f_1$. (d) Phase measurement (solid lines) and calculation (dashed lines) of the transmission parameters of the TSJPC device versus input frequency $f_1$. The phase measurement exhibits a differential phase of $180^{\circ}$ on resonance, which is a distinct feature of a gyrator. The device parameters used in the calculation are $\omega_p/2\pi=3.332$ $\operatorname{GHz}$, $\omega_{a1}/2\pi=6.948$ $\operatorname{GHz}$, $\gamma_{a1}/2\pi=60$ $\operatorname{MHz}$, $\gamma_{b1}/2\pi=55$ $\operatorname{MHz}$, $\omega_{a2}/2\pi=6.945$ $\operatorname{GHz}$, $\gamma_{a2}/2\pi=50$ $\operatorname{MHz}$, $\gamma_{b2}/2\pi=45$ $\operatorname{MHz}$, $\tau_d=0.847$ $\operatorname{ns}$, $\lvert\rho\rvert=0.9$, $\lvert\alpha\rvert=0.596$, $\varphi_1=\pi/2$, and $\varphi_2=0$.  
  		}
  		\label{AmplPhase}
  	\end{center}
  \end{figure*}
   
\section{The device}
The nonreciprocal phase-shifter scheme, used to demonstrate a gyrator operation, employs two dissipationless, nondegenerate three-wave mixers. The general scheme for the two-port, nonreciprocal phase shifter is depicted in Fig. \ref{Device} (a). It consists of two unitary frequency-conversion stages (i.e., three-wave mixers) connected together by a transmission line. Incoming signals at frequency $f_1$ on port 1 (2) of the device are upconverted using the first (second) frequency-conversion stage to frequency $f_2$ ($f_2>f_1$) and transmitted via the transmission line to the second (first) frequency-conversion stage, where it is downconverted back to $f_1$ and exits through port 2 (1). Both frequency-conversion processes taking place in the device, i.e., upconversion and downconversion, are enabled via coherent energy exchange with the pump drives feeding the two mixing stages at a certain power, and whose frequency $f_p$ corresponds to the frequency difference $f_p=f_2-f_1$. In addition to the critical role played by the pump power and frequency in the operation of the device, which will be made clear below, the pump phase plays a pivotal role as well. It controls the nonreciprocal phase shift experienced by the transmitted signals from port 1 to 2, i.e., $\varphi_d+\varphi$, versus the phase shift $\varphi_d-\varphi$ acquired in the opposite direction, where $\varphi \equiv \varphi_1-\varphi_2$ is the phase difference between the two pump drives feeding the two mixing stages and $\varphi_d$ is the phase shift introduced by the connecting components. Note that, since the nonreciprocal phase imprinted on the transmitted signals depends on the gradient of the pump phase, it does not depend on the gauge degree of freedom and thus it is gauge invariant \cite{AhranovBohmPhotonic}. In Fig. \ref{Device} (b), we exhibit a black-box representation of the device, which emphasizes the nonreciprocal phase shift introduced by the device and its dependence on the phase difference between the two pump drives feeding the system. In Fig. \ref{Device} (c), we show in more detail how the proof-of-principle device is realized, i.e., by using two separate, nominally identical Josephson parametric converters (JPCs) \cite{JPCreview}, connected through a $10.16$ $\operatorname{cm}$ (4") coaxial line. 

In general, JPCs are used as quantum-limited JPAs for qubit readout \cite{FluxoniumJumps,FeedbackJPC}, but they can also function as dissipationless three-wave mixers \cite{Conv,QuantumNode}. One main advantage of operating the JPC in the mixing mode for the gyration application instead of amplification is that, in the mixing mode, the frequencies of the transmitted signals through the device are converted without photon gain. As a result, the JPC is not required, according to Caves' theorem \cite{Caves,QuantumNoiseIntro} to add any noise to the processed signals. As seen in Fig. \ref{Device} (c), the JPC comprises a Josephson ring modulator (JRM) embedded at the center of two orthogonal, half-wavelength microstrip resonators denoted as \textit{a} and \textit{b}. The JRM consists of four Josephson junctions (JJs) arranged in a Wheatstone-bridge configuration and functions as a nonlinear, dispersive mixing element. The resonators \textit{a} and \textit{b} support two microwave tones denoted as signal (S) (at frequency $f_1$) and idler (I) (at frequency $f_2$), respectively. These tones lie within the bandwidths $\gamma_a/2\pi$ and $\gamma_b/2\pi$ of resonators \textit{a} and \textit{b}, whose fundamental modes, at frequencies $f_a=\omega_a/2\pi$ and $f_b=\omega_b/2\pi$, have an rf-current antinode at the JRM location. Both resonators are capacitively coupled to external feedlines that carry microwave signals into and out of the JPC. To address the \textit{a} and \textit{b} modes and apply the pump drive (P) to the device, both the S and I tones enter and exit through the delta ports of two off-chip, broadband $180^{\circ}$ hybrids (which define the S and I ports of the JPC), while the pump at $f_p=\omega_p/2\pi$ is fed through the sigma port of the hybrid connected to port \textit{b} (the other sigma port connected to port \textit{a} is terminated by a $50$ Ohm cold load). Furthermore, to facilitate the mixing operation in the device, a dc circulating current is induced in the outer loop by an external magnetic flux threading the JRM \cite{JPCreview}. The four large JJs inside the JRM loop serve as a linear shunt inductance for the outer JJs, which lifts the hysteretic response of the JRM versus flux and therefore makes the resonance frequencies of the JPC tunable \cite{Roch}. 

It is worth noting that the two resonance structures \textit{a} and \textit{b}, which incorporate the JRM at their center, play three important roles: (1) They set center frequencies and bandwidths for the two differential modes of the JRM, which, in turn, set the dynamical bandwidth of the device. (2) They increase the interaction time between the intraresonator photons taking part in the mixing operation and the JRM. (3) They inhibit the generation of harmonics or unwanted signals, as a result of the mixing process, whose frequencies do not lie within the bandwidths of the fundamental modes.  

\section{Noiseless Frequency Conversion}
Figure \ref{AmplPhase} (a) shows a signal-flow graph for a JPC operated in  
frequency-conversion mode. In the stiff-pump approximation, the scattering parameters of the device read \cite{JPCreview}

\begin{align}
\begin{array}
[c]{cc}%
r_{aa}=\dfrac{\chi_{a}^{-1*}\chi_{b}^{-1}-\lvert\rho\rvert^2}{\chi_{a}^{-1}\chi_{b}^{-1}+\lvert\rho\rvert^2}, &
t_{ab}=\dfrac{2i\lvert\rho\rvert e^{-i\varphi_p}}{\chi_{a}^{-1}\chi_{b}^{-1}+\lvert\rho\rvert^2}, \\
r_{bb}=\dfrac{\chi_{a}^{-1}\chi_{b}^{-1*}-\lvert\rho\rvert^2}{\chi_{a}^{-1}\chi_{b}^{-1}+\lvert\rho\rvert^2},&
t_{ba}=\dfrac{2i\lvert\rho\rvert e^{i\varphi_p}}{\chi_{a}^{-1}\chi_{b}^{-1}+\lvert\rho\rvert^2},  
\end{array}
\label{JPC_params}%
\end{align}

where $r_{aa}$ ($r_{bb}$) is the reflection parameter off port \textit{a} (\textit{b}), $t_{ba}$ ($t_{ab}$) is the transmission parameter from port \textit{a} (\textit{b}) to port \textit{b} (\textit{a}) with frequency upconversion (downconversion), and $\lvert\rho\rvert$ is a dimensionless pump amplitude. In Eq. (\ref{JPC_params}), the $\chi's$ are the bare response functions of modes \textit{a} and \textit{b} (whose inverses depend linearly on the S and I frequencies): 

\begin{align}
\chi_{a}^{-1}[\omega_{1}]=1-2i\dfrac{\omega_{1}-\omega_{a}}{\gamma_{a}}, \nonumber \\ 
\chi_{b}^{-1}[\omega_{2}]=1-2i\dfrac{\omega_{2}-\omega_{b}}{\gamma_{b}}. 
\label{Chi_params}%
\end{align}

Since the applied pump frequency satisfies the relations $\omega_{p}=\omega_{b}-\omega_{a}=\omega_{2}-\omega_{1}$, $\chi_{b}^{-1}$ of Eq. (\ref{Chi_params}) can be rewritten as  $\chi_{b}^{-1}[\omega_{1}]=1-2i(\omega_{1}-\omega_{a})/\gamma_{b}$. On resonance, the scattering parameters of Eq. (\ref{JPC_params}) can be cast in the form

\begin{align}
\left(
\begin{array}
[c]{cccc}%
\rm{cos}(\tau_0) & ie^{-i\varphi_p}\rm{sin}(\tau_0) \\
ie^{i\varphi_p}\rm{sin}(\tau_0) & \rm{cos}(\tau_0) 
\end{array}
\right),  \label{S_JPCres}%
\end{align}

using the substitution relation tanh(\textit{i}$\tau_0$/2)=$i|\rho|$. The scattering parameters of Eqs. (\ref{JPC_params}) and (\ref{S_JPCres}) embody five important properties that are crucial for understanding the device physics. They are as follows: (1) The scattering matrix is unitary (the total number of photons is conserved); i.e., the following relations hold, $\lvert r_{aa}\rvert^2+\lvert t_{ab}\rvert^2=1$ and $\lvert r_{bb}\rvert^2+\lvert t_{ba}\rvert^2=1$. (2) The transmitted signals from port \textit{a} to \textit{b} (\textit{b} to \textit{a}) undergo frequency upconversion (downconversion). (3) On resonance ($\omega_{1,2}=\omega_{a,b}$) and for a sufficiently large pump drive $|\rho|\rightarrow1^-$, the reflections off the device ports vanish ($|r|\rightarrow0$), while the transmission with frequency conversion approaches unity ($|t|\rightarrow1$). 4) The phase of the transmitted signals depends on the phase of the pump drive $\varphi_p$ [as seen in the expressions for $t_{ab}$ and $t_{ba}$ in Eq. (\ref{JPC_params})]. In general, this phase is gauge dependent since the time origin of the modulation can be ambiguous \cite{AhranovBohmPhotonic,AhranovBohmMixers}. However, in this experiment (as well as in other quantum-information-processing experiments in the microwave domain), the pump phase is well defined in reference to a common clock (i.e., the $10$ $\operatorname{MHz}$ reference oscillator of a rubidium atomic clock), which phase locks all microwave sources and measurement devices. (5) The phase shift acquired by signals transversing the device is nonreciprocal (i.e., $\varphi_p$ from port \textit{a} to \textit{b} and $-\varphi_p$ in the opposite direction).

\section{Nonreciprocal phase shifter}
The signal-flow graph for the two-stage JPC (TSJPC) device is shown in Fig. \ref{AmplPhase} (b). The scattering parameters in the graph with the subscript 1 (2) correspond to those of JPC 1 (2). The parameter $\alpha=|\alpha|e^{i\varphi_{d}}$ incorporates the amplitude attenuation coefficient $|\alpha|$ and the phase shift $\varphi_{d}=\omega_2\tau_d$ experienced by the idler signals propagating between the two JPC stages at frequency $f_2$. In the expression for $\varphi_{d}$, $\tau_d$ represents the delay time, which can, in turn, be expressed as $\tau_d=l_d\sqrt{\varepsilon}/c$, where $c$ is the speed of light, and $l_d$ and $\varepsilon$ are the effective electrical length and dielectric constant of the coaxial lines and hybrids connecting the two stages, respectively. By inspection, the scattering parameters of the two-port device can be written in the form \cite{Pozar}

\begin{align}
\begin{array}
[c]{cc}%
S_{11}=r_{aa1}+\dfrac{r_{bb2}t_{ba1}t_{ab1}\alpha^2}{1-r_{bb1}r_{bb2}\alpha^2}, & S_{12}=\dfrac{t_{ab1}t_{ba2}\alpha}{1-r_{bb1}r_{bb2}\alpha^2}, \\
S_{22}=r_{aa2}+\dfrac{r_{bb1}t_{ba2}t_{ab2}\alpha^2}{1-r_{bb1}r_{bb2}\alpha^2},&S_{21}=\dfrac{t_{ab2}t_{ba1}\alpha}{1-r_{bb1}r_{bb2}\alpha^2},
\end{array}
\label{Gyr_params}%
\end{align}

where $S_{11}$ ($S_{22}$) is the reflection parameter off port 1 (2), $S_{21}$ ($S_{12}$) is the transmission parameter from port 1 to 2 (2 to 1) (with frequency preservation). Similar to the case of one JPC, on resonance ($\omega_{1,2}=\omega_{a,b}$) and for a sufficiently large pump drive $|\rho|\rightarrow1^-$, the reflections off the device ports vanish ($S_{11},S_{22}\rightarrow0$), while the transmission amplitude approaches the attenuation amplitude set by the losses in the connecting stages ($|S_{12}|, |S_{21}|\rightarrow|\alpha|$). However, the phase acquired by the transmitted signals on resonance from port 1 to 2, i.e., $\angle S_{21}=\varphi_d+\varphi$ can be different from the phase acquired in the opposite direction, i.e., $\angle S_{12}=\varphi_d-\varphi$. While the total phase shift in each direction depends on the value of $\varphi_d$ and $\varphi$, the differential phase between the two directions depends only on $\varphi$, i.e., $\angle S_{21}-\angle S_{12}=2\varphi$. Thus, by setting the pump phase difference to $\pi/2$, the TSJPC device can serve as a gyrator, which introduces a differential phase of $\pi$ between signal beams propagating in opposite directions.    

\section{Experimental results}
\begin{figure}
  	[tb]
  	\begin{center}
  		\includegraphics[
  		width=\columnwidth 
  		]%
  		{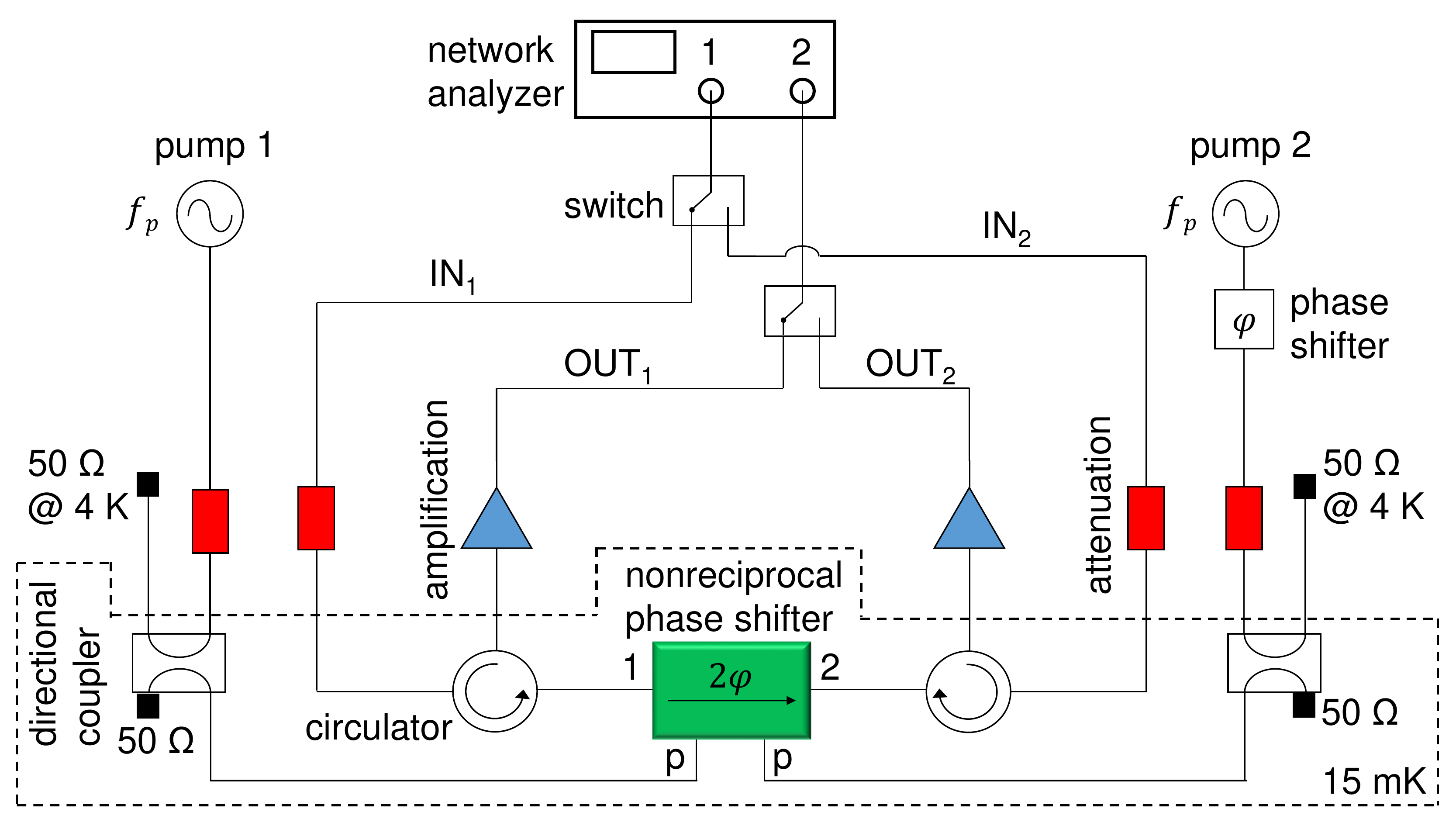}
  		\caption{A block diagram of the experimental setup used in the measurement of the scattering parameters of the TSJPC. The diagram features only the main microwave components in the setup. It does not show the distribution of the attenuation (amplification) on the input (output) lines on the different temperature stages in the dilution fridge or the pair of isolators that are installed on each output line following the circulator. It also replaces the nonreciprocal phase-shifter device shown in Fig. \ref{Device} (c) with a black-box representation. Each port of the device is connected to separate input and output lines via three-port circulators. The four scattering parameters of the device are measured using a network analyzer, whose two ports are connected to either one of the four possible pairs of input and output lines using switches located at room temperature. At the mixing-chamber stage, the incident pump powers are attenuated using $20$ dB directional couplers, which direct the unused portions of the pumps towards $50$ Ohm terminations at the $4$ K stage. The reflected pump power off the device is dissipated in a $50$ Ohm termination at the base. No additional heating of the mixing chamber is observed in this experiment due to the application of the pump drives. 
  		}
  		\label{setup}
  	\end{center}
 \end{figure}

In Figs. \ref{AmplPhase} (c) and \ref{AmplPhase} (d), we demonstrate a gyrator operation using the TSJPC device. In Fig. \ref{AmplPhase} (c), we depict using solid lines a network-analyzer magnitude measurement of the scattering parameters of the TSJPC device taken versus input frequency $f_1$. The resonance frequencies of resonators \textit{a} of the two stages are flux tuned by about $1.1\Phi_0$ to coincide at $6.945$ $\operatorname{GHz}$, where $\Phi_0=h/2e$ is flux quantum. Both JPCs are operated in frequency-conversion mode with $f_p=3.332$ $\operatorname{GHz}$. The measurement of the four scattering parameters of the device is enabled by connecting each port of the device to separate input and output lines via a three-port cryogenic circulator as illustrated in Fig. \ref{setup}. As expected, at resonance, the reflection parameters $S_{11}$ (red) and $S_{22}$ (magenta) exhibit a dip $\simeq-15$ dB, while the transmission parameters $S_{21}$ (blue) and $S_{12}$ (cyan) exhibit a peak of about $\simeq-4.6$ dB, which matches the insertion loss of the microwave components connecting the two stages at frequency $f_2$, i.e., the normal-metal coaxial cables and hybrids. Further details regarding the \textit{in situ} calibration of the amplitude of the scattering parameters are included in Appendix A. In Fig. \ref{AmplPhase} (d), we exhibit a phase-shift measurement of the transmitted signals $S_{21}$ (the solid blue line) and $S_{12}$ (solid cyan line), versus input frequency $f_1$. The phase measurement is taken at the same working point as Fig. \ref{AmplPhase} (c). The phase difference between the pump drives feeding the two JPC stages is set to yield a relative phase shift of about $180^{\circ}$ between signals transmitted in opposite directions. The dashed lines in Figs. \ref{AmplPhase} (c) and \ref{AmplPhase} (d) exhibit a theoretical calculation for the scattering parameters of the device based on Eq. (\ref{Gyr_params}) and the device parameters, which yields a relatively good agreement with the data. It is interesting that the calculation reproduces multiple unintuitive features seen in the experiment [Fig. \ref{AmplPhase} (c)], such as bends and plateaus, which can be attributed to dispersion effects of the waves propagating in the connecting microwave components and to reflection effects, which result from a certain mismatch in the relatively narrow bandwidths of the two JPCs. We also measure, at this working point, the maximum input power above which the device starts to saturate (i.e., the transmission magnitude of the device drops by $1$ dB). We find it to be around $-92$ dBm, which is comparable to the measured values in microstrip JPCs operated in frequency conversion \cite{Conv}.
  
\begin{figure}
	[tb]
	\begin{center}
		\includegraphics[
		width=\columnwidth 
		]%
		{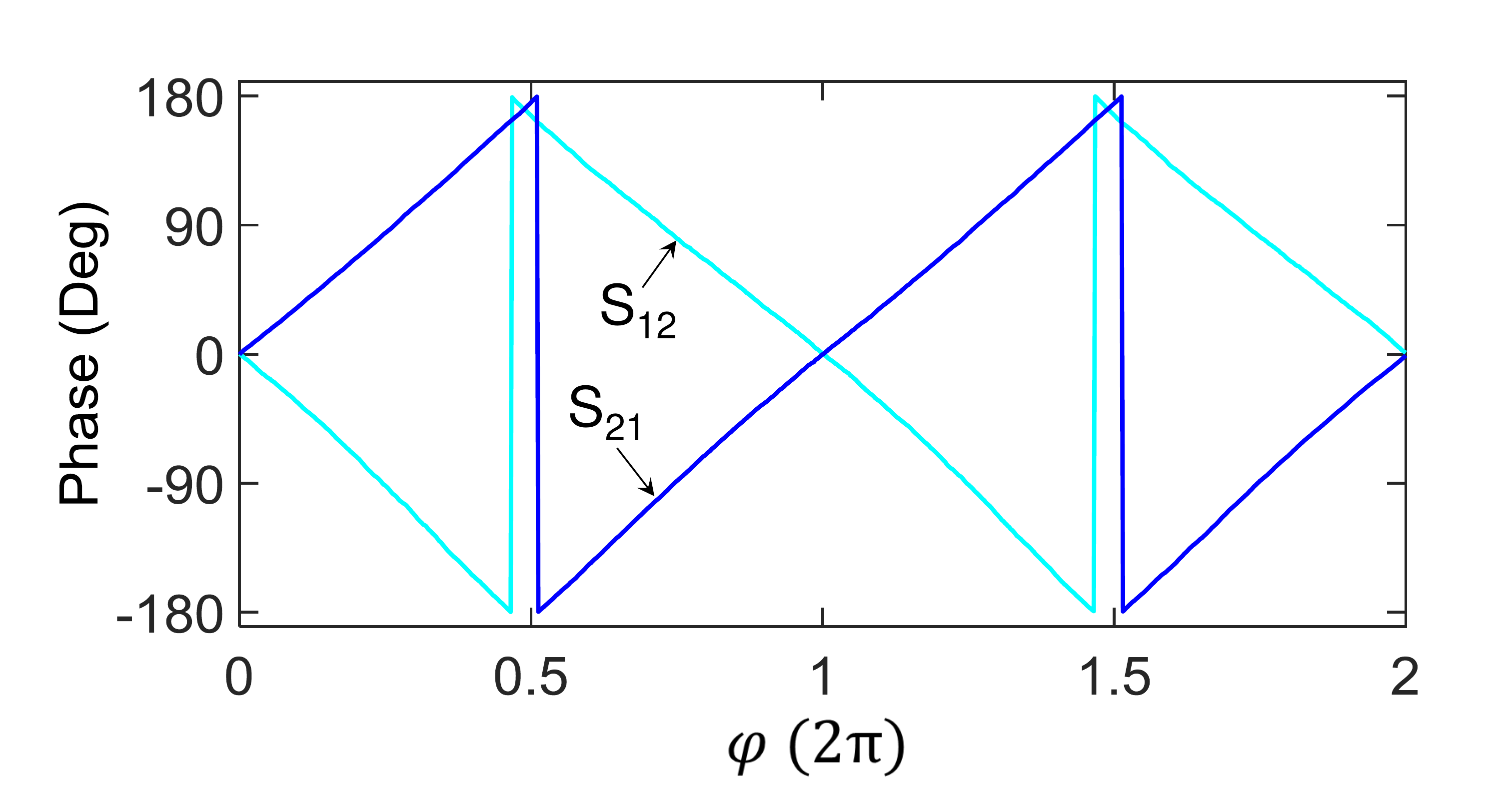}
		\caption{A network-analyzer phase measurement of the transmission parameters (i.e., $S_{21}$ and $S_{12}$) of the TSJPC device taken on resonance for varying the phase difference between the applied pump drives feeding the two stages. The device working point is the same as in Fig. \ref{AmplPhase} (c). A constant phase offset is applied to both curves in order to center them around zero phase. 
		}
		\label{NPvsPumpPh}
	\end{center}
\end{figure}

Furthermore, in Fig. \ref{NPvsPumpPh}, we display a phase-shift measurement of the transmitted signals through the device in opposite directions, taken on resonance as a function of the applied phase difference $\varphi$. As seen in the data, the phase shifts of $S_{21}$ and $S_{12}$ vary in opposite directions and depend linearly on $\varphi$, which fully agree with the theoretical prediction. This measurement result shows that the device can not only serve as a gyrator, but also as a general-purpose, nonreciprocal phase shifter, which can be fully and rapidly controlled \textit{in situ} by a microwave signal, i.e., the pump. The upper limit on such fast control is mainly set by the dynamical bandwidth of the device.

We also verify in a separate measurement (not shown) that the device at the gyrator working point does not generate any undesired harmonics in response to an input at $f_1$ or cause any observable power leakage between ports \textit{a} and \textit{b} at the idler frequency $f_2$. This is done by measuring the output signal of the device in a wideband frequency span $2-11$ $\operatorname{GHz}$, using a spectrum analyzer while inputting a signal at $f_1$. However, in this measurement, we observe a certain power leakage at the pump frequency between ports \textit{a} and \textit{b}. This is likely the case because the pump frequency $3.332$ $\operatorname{GHz}$ lies outside the band of the off-chip $180^{\circ}$ hybrids used for injecting the pumps to the device, which is $6-20$ $\operatorname{GHz}$. This undesired feature can be mitigated in the future by using hybrid-less JPCs in which the pump drive is injected through a separate physical port \cite{hybridLessJPC}.

\section{Circulator scheme}

\begin{figure}
	[tb]
	\begin{center}
		\includegraphics[
		width=\columnwidth 
		]%
		{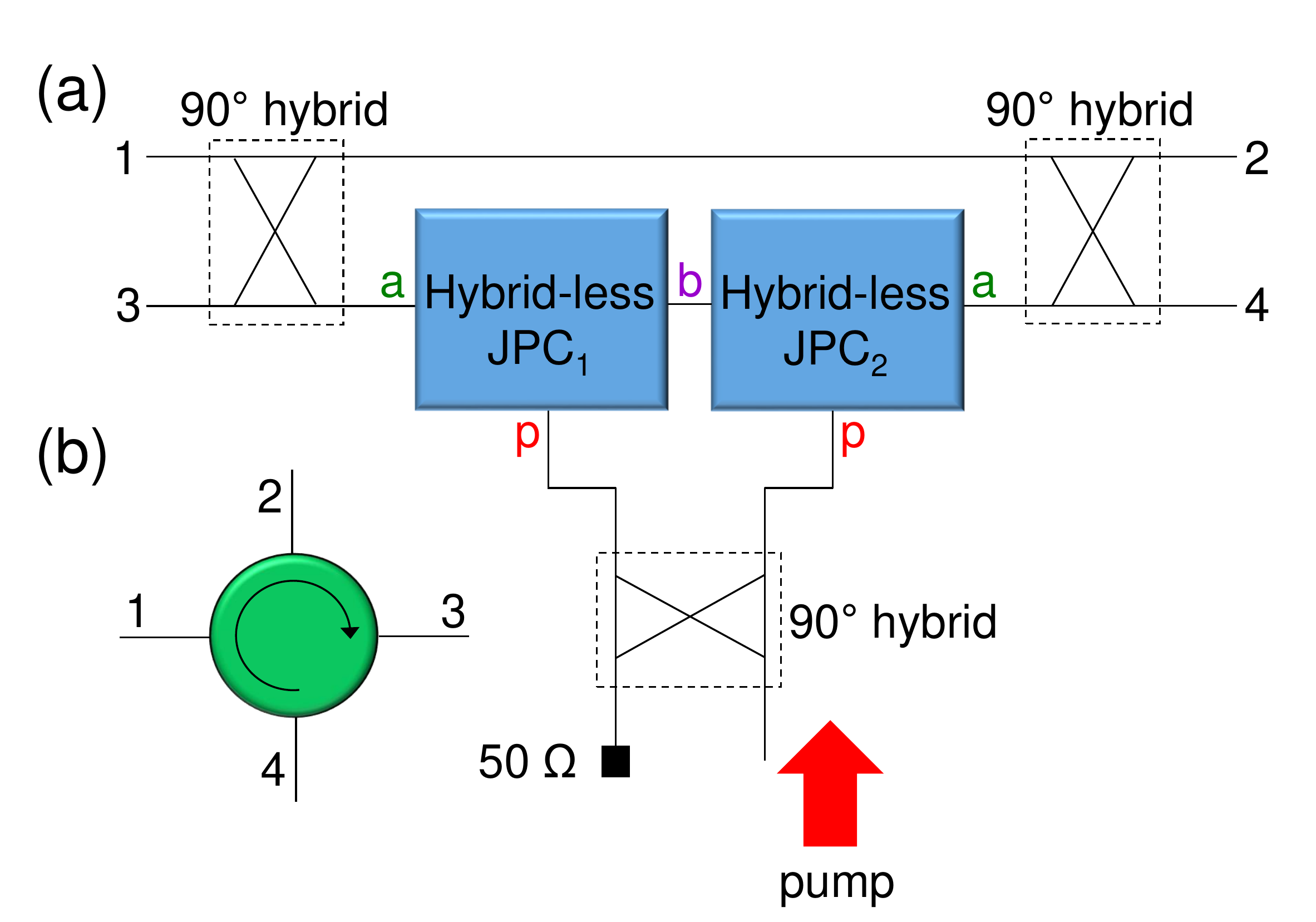}
		\caption{(a) An envisioned realization of a compact, on-chip, four-port, superconducting circulator based on the JPC-based gyrator. The circulator is formed using TSJPC coupled to three $90^{\circ}$ hybrids. The TSJPC is formed using hybrid-less JPCs with a separate spatial port for the pump drive \cite{hybridLessJPC}. Using such a design would eliminate the need for the four bulky, broadband hybrids used in this work. The JPC size can be reduced further by replacing the microstrip resonators with lumped-element capacitors shunting the JRM \cite{JPCreview,LumpedJPC}. The two $90^{\circ}$ hybrids on either side of the TSJPC are identical with a bandwidth centered around $f_1$. The bottom $90^{\circ}$ hybrid, whose bandwidth is centered around $f_p$, couples to the pump ports of the two JPCs. It enables driving the two stages using one pump tone, while simultaneously satisfying the pump phase-gradient requirement needed for the operation of the JPC-based gyrator. The three hybrids can be realized using compact coplanar waveguides or lumped-element inductors and capacitors. (b) A circuit symbol for the circulator scheme featured in (a). The arrow indicates the circulation direction between the ports, while the green color indicates that the device preserves the frequency of the transmitted signals $f_1$ across all ports.          
		}
		\label{circScheme}
	\end{center}
\end{figure}

In Fig. \ref{circScheme} (a), we present a circuit diagram of our proposed circulator scheme. It consists of (1) two hybrid-less JPCs \cite{hybridLessJPC} realized back to back on the same chip or in close proximity to each other ($\varphi_d\approx0$). By employing hybrid-less JPCs, which have separate physical ports for the pump drives, the four $180^{\circ}$ hybrids used in this work, as well as the intermediate components such as connectors and coaxial cables can be eliminated. Also, the footprint of JPCs can be significantly reduced by replacing the microstrip resonators with a lumped-element realization \cite{JPCreview,LumpedJPC}. (2) Two identical $90^{\circ}$ hybrids, which together with the TSJPC form a Mach-Zehnder interferometer, as illustrated in Fig. \ref{circScheme} (a). In this scheme, the TSJPC is incorporated into one arm of the interferometer and the hybrids have bandwidths that are centered around $f_1$. (3) One $90^{\circ}$ hybrid, whose bandwidth is centered around $f_p$, is used to deliver one pump tone to the two pump ports of the TSJPC. In addition to delivering an equal pump amplitude and frequency to the TSJPC, the $90^{\circ}$ hybrid plays another crucial role of satisfying the pump phase-gradient requirement needed for the operation of the JPC-based gyrator, namely, imposing a relative phase of $\pi/2$ between the pumps driving the two stages. The three $90^{\circ}$ hybrids can be implemented on chip using a coplanar waveguide geometry \cite{CPWhybrids} or lumped elements \cite{Lumpedhybrids}. Figure \ref{circScheme} (b) exhibits a circuit symbol for the circulator scheme introduced in Fig. \ref{circScheme} (a). Using this scheme, it is straightforward to design and implement circulators having center frequencies $f_1$ in the range $5-15$ $\operatorname{GHz}$, which matches the frequency range of working JPCs \cite{DircJPC,JPCreview,Conv,LumpedJPC}.

Since coplanar or lumped-element hybrids can have bandwidths in the range $0.5-1.5$ $\operatorname{GHz}$, the main bandwidth limitation of this scheme is set by the JPC bandwidth, which is on the order of $10$ $\operatorname{MHz}$ \cite{JPCreview,LumpedJPC}. One possible method to significantly increase this bandwidth to several hundreds of megahertz entails making the following design changes to the JPC and its coupling to the feedlines: (1) implement lumped-element JPCs \cite{JPCreview,LumpedJPC} that are analogous to lumped-element JPAs, (2) provide a separate physical feedline for the pump drive that is distinct from the signal and idler feedlines \cite{hybridLessJPC}, and (3) replace the coupling capacitors between the JPC and its signal and idler feedlines by series $LC$ circuits, whose properties are outlined for the case of JPAs in Ref. \cite{JPAimpedanceEng}.  
 
Based on the viable assumptions that (1) lumped-element JPCs and coplanar-waveguide hybrids \cite{CPWhybrids} can be compact and realized on chip, (2) the JRMs can be flux biased using on-chip flux lines, (3) the circulator chip is housed in a small, copper package, (4) the device ports consist of SMA connectors, and (5) a printed circuit board connects the SMA connectors and the circulator chip, we estimate the volume and the weight of our devices to be less than $6$ $\operatorname{cm^3}$ and $28$ $\operatorname{g}$, which are factors of $4.7$ and $1.4$ smaller than those of cryogenic, commercial circulators. Further reduction in size is possible by replacing the coplanar-waveguide hybrids with lumped-element versions \cite{Lumpedhybrids}, increasing the pump frequency, and swapping the SMA connectors with a more miniature substitute. 
 
It is important to point out that, despite these significant reductions in volume and weight, it is unlikely that, in scalable quantum architectures, these circulators would be used as stand-alone components. They are more likely to be integrated on chip or into a printed circuit board with other microwave components, which is, in turn, expected to reduce the number of connectors and yield a further reduction in size and weight.

Finally, one comment regarding the thermalization of the device: It is quite possible that, in a large quantum system, which utilizes many of these circulators, the heat generated by the pump power feeding these active circulators could exceed the cooling power of the mixing chamber. In that case, some of the reflected pump power could be routed and dissipated at higher-temperature stages (instead of the mixing chamber), which have larger cooling power, such as the still or the $4$ $\operatorname{K}$ stage. While such an arrangement would effectively increase the footprint of the setup, it can be executed in a very efficient and compact manner by integrating arrays of circulators on the same chip or the same printed circuit board and combining the reflected pumps into a small set of lines.   
    
\section{Comparison with other circulator schemes}

As outlined in the Introduction, several on-chip circulator schemes have been proposed recently and numerous proof-of-principle circulators have been demonstrated. In this section, we focus our discussion on three schemes, which share certain common features with our proposal. The first is the theoretical circulator scheme presented by Kamal \textit{et al.} \cite{NoiselessCirc}. The second and the third are the experimental works by Sliwa \textit{et al.} \cite{ReconfJJCircAmpl} and Lecocq \textit{et al.} \cite{NRAumentado2}, which demonstrate circulation using Josephson devices.  
 
The main two similarities between our circulator scheme and the one proposed in Ref. \cite{NoiselessCirc} are that both rely on two-stage, active, upconverter and downconverter Josephson devices to generate a nonreciprocal response for microwave signals, and they both break reciprocity by applying a gradient of the pump phase between the two stages (which is analogous to the role played by the magnetic field in a Faraday medium). On the other hand, the two schemes have several important differences: (1) The theoretical scheme of Ref. \cite{NoiselessCirc} is applicable to a broad group of parametric active devices. However, in practice, if realized using JPCs, it is expected to handle relatively low frequency signals ($<0.1$ $\operatorname{GHz}$). Our scheme, on the other hand, would enable circulation for microwave signals in the gigahertz range. (2) The pump frequency in Ref. \cite{NoiselessCirc} plays the role of a carrier frequency, which is modulated by the signal frequency, which in turn produces two idler frequencies. In our device, by contrast, the signal is upconverted to a single idler frequency using a pump at the frequency difference. (3) The scheme of Ref. \cite{NoiselessCirc} involves amplification and deamplification of signals. In our case, no amplification or deamplification is applied by the active devices. (4) The scheme of Ref. \cite{NoiselessCirc} requires a phase shifter between the two stages, which is not necessary in our device.   

With regard to the circulator schemes of Refs. \cite{ReconfJJCircAmpl} and \cite{NRAumentado2}, the physical mechanism for generating nonreciprocity is very similar to the one utilized here. They both rely on parametric frequency-conversion processes taking place between three resonant modes of the respective Josephson devices. Both schemes also use the relative phase of the pumps driving the Josephson devices in order to induce nonreciprocal response, i.e., circulation. The main two differences, however, between our scheme and those of Refs. \cite{ReconfJJCircAmpl,NRAumentado2} lie in the devised coupling between the three modes and how wave interference is generated. While the schemes of Refs. \cite{ReconfJJCircAmpl,NRAumentado2} employ three distinct-frequency resonance modes of the same Josephson device, ours employs two identical pairs of resonance modes existing in two separate Josephson devices (each having modes \textit{a} and \textit{b}), in which one pair with the same frequency (e.g., two \textit{b} modes) are coupled together, thus serving as an internal mode of the system. Also, while the schemes of Refs. \cite{ReconfJJCircAmpl,NRAumentado2} interfere with frequency-converted signals internally, inside the Josephson device, our scheme generates interference externally, using the Mach-Zehnder interferometer. 

Some potential advantages of our proposed circulator scheme over the circulators presented in Refs. \cite{ReconfJJCircAmpl,NRAumentado2} include the following: 

(1) It realizes circulators with four ports versus three. The additional port can be used to enhance connectivity in quantum networks, reduce the total number of circulators in quantum processors, and improve thermal anchoring (e.g., by connecting one port of the circulator to a cold termination). As an example for enhanced connectivity using the four-port circulator, an input line is connected to port 1, a quantum-resonator system to port 2, a JPA (working in reflection) to port 3, and an output line to port 4. 

(2) It preserves the frequency of the processed signal across all ports. This property can be useful, for example, in reducing the number of microwave generators required for qubit readout. In state-of-the-art qubit readout schemes, in which the frequency of the readout is preserved, one generator is sufficient. In such a setup, the output signal of the generator is split into two parts. The frequency of one portion, serving as the readout signal for the qubit-resonator system, is offset by tens of megahertz (e.g., $\sim 10$ $\operatorname{MHz}$) using an in-phase quadrature mixer, which, in turn, is modulated by an arbitrary-wave-generator control signal. Such control signals are required in qubit measurements in order to orchestrate the timing, shape, and sequence of the various qubit and readout pulses. The other portion of the generator signal facilitates the downconversion of the gigahertz-range readout signal exiting the fridge down to a megahertz-range signal ($\sim 10$ $\operatorname{MHz}$) using a room-temperature mixer. It is this downconverted signal that is ultimately digitized and measured in the majority of qubit-readout experiments. However, when using a frequency-converting circulator for readout such as those in Refs. \cite{ReconfJJCircAmpl,NRAumentado2}, two microwave generators would be necessary because the input and output readout signals are several gigahertz apart. 

(3) It requires, for its operation, one pump tone versus three. This can result in a significant reduction by a factor of three in the number of microwave generators needed for the operation of superconducting circulators in large quantum networks. 

(4) It has four degrees of freedom/control knobs needed for its operation (i.e., the fluxes threading the two JRMs, the amplitude and frequency of one microwave drive) versus seven (i.e., the flux threading the JRM/dc-SQUID, the amplitudes of the three pumps, the algebraic sum of their phases, and the frequencies of two out of the three pumps). Such a reduced number of degrees of freedom is expected to enhance the device stability over time and require simpler control and feedback schemes. 

(5) It allows for a partial transmission of microwave signals between its ports, which the circulator is not specifically designed to block or route (e.g., out-of-band signals). This can be particularly useful when certain out-of-band signals serve only as inputs for a single-port quantum system. One example of such a scenario is a superconducting qubit dispersively coupled to a readout resonator measured in reflection. In this example, both the readout and qubit pulses, which can be a few gigahertz apart, can be applied to the narrowband circulator and feed the single-port qubit-resonator system connected to one port of the circulator. Assuming that the readout (qubit) signal of the qubit-resonator lies in band (out of band) of the circulator, the readout signal would be strictly routed between the circulator ports based on the circulation direction, while the qubit signal would reach all four ports equally. 

\section{Summary}

In this paper, we implement and measure a proof-of-principle, nonreciprocal phase shifter that does not employ any magnetic materials or strong magnets. The device is realized by coupling two dissipationless, nondegenerate Josephson mixers. By operating the Josephson mixers in frequency-conversion mode, we show that the phase shift acquired by signals transmitted through the device can be controlled \textit{in situ} by the phase difference of the pump drives feeding the two mixers. We also show that by setting the phase difference of the pump drives to $\pi/2$, the device can operate as a microwave gyrator. Both results are found to be in good agreement with the device theory.

Looking forward, the performance and size of such a nonreciprocal phase shifter can be significantly improved by eliminating the loss and delay between the two mixing stages. This can be achieved, for example, by integrating the two Josephson mixers on the same chip (or printed circuit board) in close proximity to each other and eliminating the need for bulky, off-chip, broadband hybrids by using a hybrid-less version of the Josephson mixers \cite{hybridLessJPC}. Also, it is feasible to significantly enhance the instantaneous bandwidth of the device to more than $600$ $\operatorname{MHz}$ by utilizing impedance-engineering techniques \cite{JPAimpedanceEng}. Such a lossless, nonreciprocal phase shifter and gyrator, which does not employ any magnetic materials or strong magnets for its operation, could be used in a variety of quantum-information-processing applications. These applications could range from \textit{in situ} manipulation of microwave signals in superconducting circuits to the realization of on-chip circulators for routing signals in a quantum processor. 

\begin{acknowledgments}
B.A. thanks Michel Devoret, Archana Kamal, and Michael Hatridge for discussions, and Christian Baks for the design and supply of printed circuit boards used in mounting the JPC devices.  
\end{acknowledgments}

\appendix

\section{\textit{In situ} calibration of the device scattering parameters}
 In order to quantify the insertion loss of the intermediate stages that separate the two JPC stages, i.e., the two hybrids and coaxial cables [see Fig. \ref{Device} (c)], we utilize three properties of the JPC operated in conversion mode: (1) Without an applied pump, the JPC acts as a perfect mirror with unity reflection off the signal (\textit{a}) and idler (\textit{b}) ports. This can be verified by comparing the on-resonance and off-resonance reflection parameters of the JPC. In general, the off-resonance level of the reflection parameters of the JPC sets the $0$ dB reference for the scattering parameters measurement. (2) With an applied pump driving the JPC in full conversion, the amplitude of the transmission between the ports is unity. This can be inferred by measuring the reflection parameters on resonance which approach zero. (3) The amplitude of the transmission parameter is equal in both directions (i.e., from port \textit{a} to \textit{b} versus \textit{b} to \textit{a}). This property implies that, on resonance, the amplitudes of the transmission parameters through the whole device must be equal (since the hybrids and coaxial cables separating the two JPCs are passive and reciprocal). We also measure, using the setup exhibited in Fig. \ref{setup}, the amplitude of the scattering parameters of the device, i.e., $|S_{11}|, |S_{22}|, |S_{21}|, |S_{12}|$. Specifically, we measure the amplitude of the reflection parameters on resonance with no applied pump (the \textit{off} state), $|S^{\rm{off}}_{11}(f_a)|$ and $|S^{\rm{off}}_{22}(f_a)|$, and the amplitude of the transmission parameters on resonance with the TSJPC operated in full conversion (the \textit{on} state), $|S^{\rm{on}}_{21}(f_a)|$ and $|S^{\rm{on}}_{12}(f_a)|$.
 
 To demonstrate how these measured parameters, combined with the JPC properties, can be used to determine the relative offset in decibels between the amplitude of the maximum transmission and the reflection parameters on resonance, we express them in the form 
  
\begin{align}
|S^{\rm{off}}_{11}(f_a)|=|O_1|-|I_1|-|L_1|,
\label{S11_calib}%
\end{align}

\begin{align}
|S^{\rm{off}}_{22}(f_a)|=|O_2|-|I_2|-|L_1|, 
\label{S22_calib}%
\end{align}

\begin{align}
|S^{\rm{on}}_{21}(f_a)|=|O_2|-|I_1|-|L_1|-|L_2|, 
\label{S21_calib}%
\end{align}

\begin{align}
|S^{\rm{on}}_{12}(f_a)|=|O_1|-|I_2|-|L_1|-|L_2|, 
\label{S12_calib}%
\end{align}

where $|I_{1}|$ ($|I_{2}|$) is the total attenuation in decibels of the input line that extends from port $1$ of the network analyzer to port $1$ ($2$) of the TSJPC at $f_a$ (see Fig. \ref{setup}), $|O_{1}|$ ($|O_{2}|$) is the net gain in decibels of the output line that extends from port $1$ ($2$) of the TSJPC to port $2$ of the network analyzer at $f_a$ (see Fig. \ref{setup}), $|L_1|$ is the insertion loss experienced by signals at $f_a$ that are either passing twice through the same hybrid and short, phase-matched coaxial cables connected to port \textit{a} of the JPC [see Fig. \ref{Device} (c)] due to reflection or passing through two identical hybrids and short, phase-matched coax cables (connected to port \textit{a} of the JPC) due to transmission through the TSJPC [see Fig. \ref{Device} (c)], and, finally, $|L_2|$ is the insertion loss in decibels of the intermediate stages in the TSJPC at $f_b$ [the two hybrids, the short, phase-matched coaxial cables, and the normal-metal coaxial cable connecting the two stages, as illustrated in Fig. \ref{Device} (c)]. 

One important observation from Eqs. (\ref{S11_calib})-(\ref{S12_calib}) is that, although the hybrid and short, phase-matched coaxial-cables connected to port \textit{a} are part of the JPC circuit, the insertion loss associated with them, $|L_1|$, cannot be distinguished (through measurements of the scattering parameters of the device only) from $|I_{1}|$ and $|O_{1}|$ on port $1$ or $|I_{2}|$ and $|O_{2}|$ on port $2$. In other words, $|L_1|$ can be lumped into $|I_{1,2}|$ and $|O_{1,2}|$ in a similar fashion to what is generally done for the loss associated with the circulators connected to the JPC ports, whose role is to separate the incoming and outgoing signals propagating on the JPC feedlines. 

This is, however, not the case for the insertion loss associated with the intermediate components between the two stages $|L_2|$. It can be evaluated by summing Eqs. (\ref{S21_calib}) and (\ref{S12_calib}) and substituting Eqs. (\ref{S11_calib}) and (\ref{S22_calib}) into the sum, which yields     

\begin{align}
|L_2|=\dfrac{1}{2}(|S^{\rm{off}}_{11}|+|S^{\rm{off}}_{22}|-|S^{\rm{on}}_{21}|-|S^{\rm{on}}_{21}|). 
\label{S_calib}%
\end{align}

In our case, we find that $|L_2|\backsimeq4.6$ dB, which is consistent with room-temperature measurements of the intermediate components connecting the two JPCs taken at $10$ $\operatorname{GHz}$ and below the loss limit of their specifications.

\end{document}